\documentclass[acmsmall]{acmart}

\pdfoutput=1

\makeatletter                   
\def\mdseries@tt{m}             
\makeatother                    
\usepackage[plain]{fancyref}
\usepackage[draft=true]{minted} 
\usepackage{color}
\usepackage{hyperref}           
\hypersetup{
    colorlinks=true,
    linkcolor=blue,
    filecolor=red,      
    urlcolor=magenta,
    breaklinks=true,            
}
\usepackage{breakurl}           

\usepackage{subfigure}
\usepackage{graphicx,epsfig}
\usepackage{algorithm2e}

\newcommand{\solution}{\textsc{\small TripDecoder}}     
\newtheorem{defn}{Definition}[section]

\AtBeginDocument{%
  \providecommand\BibTeX{{%
    \normalfont B\kern-0.5em{\scshape i\kern-0.25em b}\kern-0.8em\TeX}}}

\begin{document}

\title{TRIPDECODER: Study Travel Time Attributes and Route  Preferences of Metro Systems from Smart Card Data}

\author{Xiancai TIAN}
\email{shawntian@smu.edu.sg}
\author{Baihua ZHENG}
\email{bhzheng@smu.edu.sg}
\author{Yazhe WANG}
\email{yzwang@smu.edu.sg}
\affiliation{%
  \institution{Living Analytics Research Centre, Singapore Management University}
  \country{Singapore}
}

\author{Hsiao-Ting HUANG}
\email{q36064248@gs.ncku.edu.tw}
\affiliation{%
 \institution{Department of Electrical Engineering, National Cheng Kung University}
 \country{Taiwan}}

\author{Chih-Chieh HUNG}
\email{smalloshin@gms.tku.edu.tw}
\authornote{The corresponding author.}
\affiliation{%
 \institution{Department of Computer Science and Information EngineeringIn, Tamkang University}
 \country{Taiwan}}


\begin{abstract}
%
In this paper, we target at recovering the exact routes taken by commuters inside a metro system that are not captured by an Automated Fare Collection (AFC) system and hence remain unknown. We strategically propose two inference tasks to handle the recovering, one to infer the travel time of each travel link that contributes to the total duration of any trip inside a metro network and the other to infer the route preferences based on historical trip records and the travel time of each travel link inferred in the previous inference task. As these two inference tasks have interrelationship, most of existing works perform these two tasks simultaneously. However, our solution \solution\ adopts a totally different approach. To the best of our knowledge, \solution\ is the first model that points out and fully utilizes the fact that there are some trips inside a metro system with only one practical route available. It strategically decouples these two inference tasks by only taking those trip records with only one practical route as the input for the first inference task of travel time and feeding the inferred travel time to the second inference task as an additional input which not only improves the accuracy but also effectively reduces the complexity of both inference tasks. Two case studies have been performed based on the city-scale real trip records captured by the AFC systems in Singapore and Taipei to compare the accuracy and efficiency of \solution\ and its competitors. As expected, \solution\ has achieved the best accuracy in both datasets, and it also demonstrates its superior efficiency and scalability. 
\end{abstract}




\maketitle

\section{Introduction}
\label{sec:introduction}

For land-scarce and metro-rely countries like Singapore, it is extremely important to improve the public transport systems in order to meet the increasing travel demands of a growing economy and population. \emph{Mass Rapid Transit} (MRT) system is a critical part of the public transport system because of its advantages in both capacity and efficiency\footnote{In this paper, the term MRT system is used interchangeably with metro system.}. In order to increase the MRT ridership and encourage more commuters to take MRT, it is critical to improve the MRT services which has attracted attention from the academy.   

For instance, predicting vehicle crowdedness and platform commuter intensity can help operators evaluate service quality and design structural improvements for the metro network~\cite{vandewiele2017,nuzzolo2013,bofriis2014}; understanding commuters' route choice preferences and route travel time allows operators to provide more accurate route recommendation~\cite{holleczek2015,jin2017}; studying commuters' movement during MRT disruption enables operators to identify potentially overcrowded stations and to take more targeted remedial actions like arranging alternative transportation modes~\cite{Silva2015,Yin2016,shawn2018}.

Studies on commuters' behaviour inside public transport systems have long been relying on external data source like field surveys~\cite{jin2017,Silva2015,rickwood2009urban,chu2010augmenting} and crowdsourcing~\cite{vandewiele2017}. However, these data sources have their own limitations. Take survey data as an example. It is easily subject to bias and errors, and conducting surveys and processing the data can be both time-consuming and labor-intensive. In addition, since most surveys are conducted with focus on particular location and time, the results are often limited in scale and diversity. Data collected via crowdsourcing suffers from similar issues. As a result, alternative data sources are required to be able to more accurately and more comprehensively understand the spatial-temporal characteristics of travel patterns, such as train control sensors~\cite{nuzzolo2013,bofriis2014}, and GPS data~\cite{holleczek2015}.

In this paper, we aim at inferring the travel time required by any route inside the metro network, and route preferences at both aggregation level and individual level, based on data collected from automated fare collection (AFC) systems that have emerged and widely deployed over the last decade. When the context is clear, we may use the term \emph{AFC data} interchangeably with the term \emph{smart card data}, \emph{trip records} or \emph{trip observations}, and they all refer to some key information related to trips (e.g., the time stamp and the  MRT station/bus stop when a trip is started, and the time stamp and the  MRT station/bus stop when a trip is ended). 

As more and more public transportation systems are now using smart cards to collect trip fares, it has generated massive precious data resource for public transport scientific study. However, the smart card data has limitations. For example, because of the reliability requirement of a metro network, redundant design is adopted to tolerate faults. Consequently, there could be multiple routes available to bring a commuter from the boarding station to the alighting station. However, as most metro networks are designed as closed systems and commuters only leave traces at boarding/alighting stations for the purpose of fare collection, the exact route taken by each individual commuter remains unknown. On the other hand, the information of each commuter's movement inside the metro network is critical to the study of commuter behaviors at a microscopic level.

As mentioned above, the main objective of this paper is to infer the travel time of any route, and to infer the route preferences of commuters if there are multiple routes available to bring the commuter from the boarding station to the alighting station. These two inference tasks have interrelationship, and hence existing works on similar topics perform these two inference tasks simultaneously, which significantly increases the complexity of the problem. We adopt a very different approach. Our solution, \solution, takes in a static metro network and its smart card data as inputs. By carefully studying the data, \solution\ points out a fact that some trips inside a metro system have \emph{only one} practical route. It makes full use of this finding, and decouples the two inference tasks into two separated steps.

During the data pre-processing stage, a route candidate set is generated for each Origin-Destination (OD) pair of stations, where unrealistic routes, such as routes that are extremely long with loops, are removed. We then category OD pairs into two disjoint sets based on the number of available routes linking them, i.e., OD pairs with a single route and OD pairs with multiple alternative routes. The clever separation of OD pairs with single route from those with multiple routes actually motivates the design of our first inference task. Accordingly, \solution\ strategically decomposes the travel time required by a trip into different travel links, and fully utilizes single-route OD pairs and their corresponding trips (captured by the AFC system) to derive travel time of different travel links that contribute to the travel time of any trip. Because 
\solution\ only considers the trips of OD pairs with single routes, there is no ambiguity in terms of the routes taken to complete the trips. Therefore, we effectively remove the dependency of the route preference from the inferring of travel time, and are able to produce more accurate estimation of the travel time of travel links. The inferred travel time of different travel links are then used to construct travel time of routes on multi-route OD pairs which becomes an additional and useful input for the inference of route preferences. With route travel time known, the complexity of the inference of the route preferences w.r.t. multiple routes has been effectively reduced.

To illustrate and verify the proposed solution, we carry out case studies using real datasets, i.e., the city scale real trip data captured by AFC systems in Singapore and Taipei. Our result demonstrates the superior performance of \solution, in terms of both accuracy and efficiency.  

The remainder of this paper is organized as follows.  In Section~\ref{sec:literature_review}, we review previous studies on several related topics, including metro network travel time estimation, commuter route choice behaviour, and the use of smart card data in understanding metro operation and flow assignment. In Section~\ref{sec:modeling_framework}, we present the preliminaries of \solution, including the formulation of the problem studied in this paper, the route choice set extraction, and data exploration insights. In Section~\ref{sec:solution_algorithm}, we present the framework of \solution\ and detail the two-step solution algorithm to recover the route travel time and to learn the route preferences. In Section~\ref{sec:case_study}, we apply \solution\ on real trip data collected from Singapore and Taipei as two case studies and report the performance of \solution. We close the paper with conclusion and discussion of future research directions in Section~\ref{sec:conclustion}. Note that without the loss of generality, in the rest of paper we use Singapore metro network and its smart card data collected during morning peak hours in 2015 December as an example to explain how the proposed framework works.

\section{Literature Review}
\label{sec:literature_review}

Understanding the commuter flow in a transportation system is an important research topic. In the existing literature, many of the works focus on studying commuter flow models based on experience~\cite{nakayama2000route,model2,model3}. The models depend heavily on behavior assumptions and hence lack reliable empirical data verification. Other studies are based on field surveys~\cite{jin2017,Silva2015,rickwood2009urban,chu2010augmenting}, crowdsourcing~\cite{vandewiele2017}, train control sensors~\cite{nuzzolo2013,bofriis2014}, and GPS data~\cite{holleczek2015}. These datasets are usually expensive to obtain, small in scale, and poor in accuracy, therefore would greatly affect the analytic power of the applications built based on them. 

In recent years, smart card data have provided us with new opportunities to perform data-centric transit behavior study. \cite{Hong2017} develops a heuristic method to assign commuter flows inside a metro network based on AFC data. The main idea is to use train timetable to estimate the pure travel time of every trip record, and then to cluster the trips based on the pure travel time between an OD pair, with the assumption that each trip cluster corresponds to a candidate route connecting the OD pair. The method is very efficient but it requires additional information of real-time train timetable, which is not always available. It also has accuracy issue due to the many assumptions made such as train services strictly follow the timetable, and commuters never fail to board on the immediate train after entering the stations. \cite{Sun2017} studies the latent relationships among OD pairs, candidate routes and commuter travel time, and obtains the distribution of commuter flow on different candidate routes by a \emph{Latent Dirichlet Allocation} (LDA) model. However, their model is not able to capture the travel time distribution on different routes, and thus could not infer local commuter flow of individual station/link segment of the metro network.

To fully exploit the AFC data and predict the local commuter flow of individual link/station and commuters' route preferences at the same time, many recent researches rely on statistical modelling based inferences~\cite{Sun2015,Sun2015-2,Xu2018,NIPS2017}. These models take commuter travel time as observations and characterize them as a mixture distribution from all potential routes. \cite{Sun2015,Sun2015-2} propose to construct posterior probability by combining the likelihood of observed commuter travel times provided by AFC data and prior knowledge about the studied transportation network. They assume the link travel time of the transit network follows the normal distribution, and the commuters' route choice probability could be represented by a logit model of various influential factors (i.e., in-vehicle travel time, transfer time). Thereafter, they perform Bayesian inference to calibrate the parameters (i.e., mean/variance of link travel time and coefficient of  influence factors of the route choice probability) of the model. \cite{Xu2018} builds a similar posterior probability model as~\cite{Sun2015,Sun2015-2}, where they cnosider not only in-vehicle travel time and transfer time factors, but also crowdedness factor for route choice probability. 

To the best of our knowledge, work presented in \cite{NIPS2017} represents the state-of-the-art solution to the inference of travel time and route preferences of commuters inside a metro network based on AFC data. It proposes a different likelihood model of the observed commuter travel time by modeling the path travel time as complicated convolutions of Poisson distributions, and models the path choice probability as a logit model of the station number factor and the transfer number factor. Due to the intractability of the model, \cite{NIPS2017} also proposes approximate inference schemes to estimate the model parameters. The models discussed above assume the commuter route choice is determined by a few predetermined influential factors (e.g., route travel time, transfer number). However, factors that affect commuters' route-choice decisions could be complicated and difficult to model, missing key influential factors may affect the accuracy of the model. 

Different from existing solutions, we adopt a data-driven approach. \solution\ models the route preferences purely based on real travel time observations reflected by the smart card data but not any explicit influential factor. In addition, there is interrelationship between these two inference tasks and all the existing works perform these two inference tasks simultaneously, which significantly increases the complexity of proposed models. The models search for the optimal parameter combination in an extremely large search space, which results in low accuracy and poor efficiency and scalability. \solution\ is designed to address both the accuracy issue and the performance issue, as we strongly believe that an ideal solution shall be able to achieve a high accuracy and to complete the inference tasks efficiently and meanwhile is scalable. To our best knowledge, this is the first work on learning the travel time and route preferences from AFC data that considers the efficiency and scalability of the inference model, in addition to the accuracy. 

A preliminary work was published in~\cite{yazhe2018}. As compared with previous preliminary work, we have made following new contributions in this extended version. First, we have improved the inference models which helps to further improve the accuracy and the efficiency of the proposed model framework. For example, we notice the entry walking time and the exit walking time may follow different distributions, as the exiting action happens right after a train reaches the station while the entering action could happen any time. Accordingly, in this extended version, we assume they follow different distributions which does improve the accuracy. Instead of modeling the travel time from station $s_i$ to its adjacent station $s_j$ and the time from $s_j$ to $s_i$ differently, we simplify the model by assuming the travel time is independent of the direction. It does help reduce the complexity of the model and hence improve the efficiency, without downgrading the accuracy. When we perform the inference, we explore the impact of initial values on the performance and the new initial values used in this extended version actually are more appropriate as the training time has been reduced. Second, we have significantly improved the experimental study. To be more specific, we have included the work published in~\cite{NIPS2017} as a new competitor; we have included the AFC data collected from Taipei as an additional dataset and reported the performance of \solution\ and its competitors based on Taipei dataset; we have designed and implemented an evaluation framework for the inference of route preference and reported the performance of \solution\ and its competitors; and we have included a new set of experiments to demonstrate the advantage of \solution\ in terms of efficiency and scalability, as compared with its competitors. Third, we have detailed the insights we have obtained from our initial study on Singapore dataset, which suggest a simple but very novel and effective approach to perform the inference of travel time. Fourth, we have significantly improved the presentation and the organization of the paper. In brief, we believe this extended version has included sufficient fresh contributions.

\section{Preliminary}
\label{sec:modeling_framework}

Before we present \solution, we first propose a trip reconstruction process in Section~\ref{subsec:problem-formulation}, which defines a trip as a sequence of steps to ease the inference of the travel time required. We formulate the metro system as a general graph network and introduce the notations used throughout the paper. Next, we introduce the concept of \emph{candidate route set} $R_{od}$ that is defined for a given OD pair $\langle o,d\rangle$ in Section~\ref{subsec:candidate_route}. We use this concept to cluster all the OD pairs into two disjoint categories, the one with only one candidate route and the other with multiple candidate routes. We then perform data exploration in Section~\ref{subsec:insight}, using AFC data collected from Singapore and Taipei, and report our findings, which lay the foundation for \solution. Table~\ref{tab:symbol} lists the symbols that will be frequently used in the rest of the paper. 

\begin{table*}[t!]
\centering
\normalsize
\caption{Frequent Symbols}
\label{tab:symbol}
\begin{tabular}{|p{1.5cm}|p{11cm}|}
\hline
\textbf{Symbol}           & \textbf{Definition} \\ \hline
$G(S,E,L)$  & a general transportation graph with $S$, $E$, $L$ representing stations, edges, and service lines respectively \\ \hline
$r_{ij}$ & a route from station $s_i$ to station $s_j$ \\ \hline
$r_{ij}.k$ & the number of links travelled by a route $r_{ij}$\\ \hline
$r_{ij}.q$ & the number of transfers required by a route $r_{ij}$\\ \hline
$|r_{ij}|$ & the length of route $r_{ij}$ which is defined as $r_{ij}.k+\alpha \times r_{ij}.q$\\ \hline
$T_{r_{ij}}$ & the travel time required by a route $r_{ij}$ \\ \hline
$T^g_s$ & entry walking time from the turnstile to the platform at station $s$ \\ \hline
$T^w_l$ & waiting time for the service line $l$ at the platform \\ \hline
$T^c_{e}$ & train travel time corresponding to an link $e$ \\ \hline
$T^q_s$ & transfer time required at interchange station $s$ \\ \hline
$T^a_s$ & exit walking time from the platform to the turnstile at station $s$ \\ \hline
$tr$  & a trip record captured by AFC data, in the form of $(id, s_o, s_d,t)$ \\ \hline
$TR_{od}$ & the set of observed trips corresponding to a given OD pair $\langle o, d \rangle$, $TR_{od} = \{tr|tr.s_o=o \land tr.s_d=d\}$ \\ \hline
$R_{od}$ &the set of routes corresponding to a given OD pair $\langle o, d \rangle$, i.e., $R_{od} = \cup r_{od}$ \\ \hline
$r^{min}_{od}$ & the route corresponding to OD pair $\langle o, d \rangle$ with the shortest length \\ \hline
$OD_s$/$OD_m$ & the set of OD pairs that have one route/multiple routes \\ \hline
$TR_s$/$TR_m$ & the set of trip observations that corresponding to the OD pairs preserved by $OD_s$/$OD_m$ \\ \hline
\end{tabular}
\vspace{-0.15in}
\end{table*}

\subsection{Problem Formulation }
\label{subsec:problem-formulation}

In this paper, we model a metro network as a general transportation graph $G(S, E, L)$, consisting of a set of metro stations $S$, a set of edges $E$, and a set of metro lines $L$. A station $s \in S$ could be either a normal station that is crossed by only one metro line or an interchange that is crossed by multiple metro lines. An edge (or a link, interchangeably) $e(s_i,s_j,l) \in E$ is a segment on a train line $l_x \in L$ that connects two stations $s_i$ and $s_j$ without passing any other station. Stations $s_i$ and $s_j$ are so called adjacent if there is an edge $e(s_i,s_j,l) \in E$ between them. Note that there could be multiple edges corresponding to two adjacent stations $(s_i, s_j)$, corresponding to different lines. 
In addition, we model a metro network as an undirected graph for simplicity. However, the techniques developed in this paper could be easily extended to support the case where a metro network is modelled as a directed graph. 

\begin{figure}[t!]
\centering
\includegraphics[width=8cm]{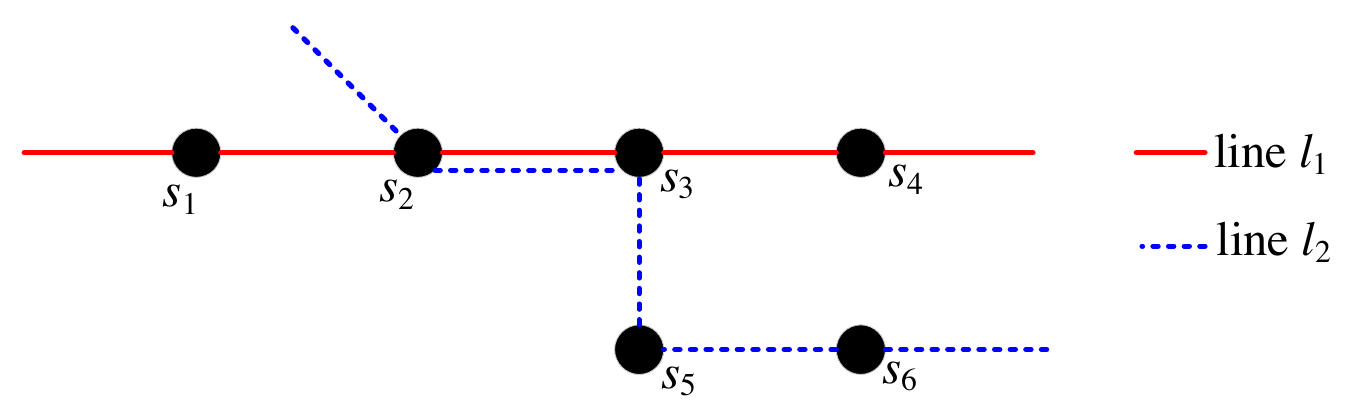}
\caption{Example network}
\label{fig:example-network}
\end{figure}

An example metro network is depicted in Figure~\ref{fig:example-network} for illustration purpose. Accordingly, we have $S=\{s_1, s_2, s_3, s_4, s_5, s_6,\cdots\}$, $E=\{e_1(s_1,s_2,l_1)$, $e_2(s_2,s_3,l_1)$, $e_3(s_3,s_4,l_1)$, $e_4(s_2,s_3,l_2)$, $e_5(s_3,s_5,l_2)$, $e_5(s_5,s_6,l_2)$ $\cdots\}$, and $L=\{l_1, l_2, \cdots\}$. Station $s_2$ and station $s_3$ are adjacent, and they are connected by two edges, i.e., $e_2$ and $e_4$ corresponding to lines $l_1$ and $l_2$ respectively. 
%
%
Stations $s_2$ and $s_3$ are also interchanges as commuters can switch from one service line to another at both $s_2$ and $s_3$, while stations $s_1$, $s_4$, $s_5$ and $s_6$ are normal stations. 

A route $r_{ij}$ from an origin station $s_i$ to a destination station $s_j$ is a sequence of adjacent edges $\langle e_1, \cdots, e_k\rangle$ that could bring commuters from station $s_i$ to station $s_j$. Edge $e_1(s_{i_1},s_{j_1},l_1)$ and edge $e_2(s_{i_2},s_{j_2},l_2)$ are adjacent if $e_1.s_{j_1}=e_2.s_{i_2}$, 
while they do not necessarily correspond to the same line. If two edges are in different line, that is $e_1.l_1 \ne e_2.l_2$, it indicates the travel from $e_1$ to $e_2$ requires a transfer from line $e_1.l_1$ to another line $e_2.l_2$ at the station $e_1.s_{j_1}$. For example, edges $e_1$ and $e_2$ are adjacent but not edges $e_1$ and $e_3$. Route $r_{15} = \langle e_1,e_2,e_5\rangle$ provides an example route from station $s_1$ to station $s_5$ which requires a transfer at station $s_3$; and  $r'_{15} =\langle e_1,e_4,e_5\rangle$ is another route from station $s_1$ to $s_5$ which requires a transfer at station $s_2$. 

In this paper, we only consider simple routes without loop, so that each route only visits a station at most once. If we assume that there is no significant difference among the travel time required by each edge, the length of a route $r_{ij}$ is determined by two parameters, the number of edges travelled and the number of transfers required, denoted by $r_{ij}.k$ and $r_{ij}.q$, respectively. Take the example route $r_{15}$ as an example. We have $r_{15}.k=3$ as it passes three edges and $r_{15}.q=1$ as it requires one transfer at the transfer station $s_3$. Since a route may include transfer stations, we generalize the length of a route by taking the number of transfers into account. Given a route $r_{ij}$, the length of $r_{ij}$ is defined as $|r_{ij}|=r_{ij}.k+\alpha\times r_{ij}.q$, e.g., $|r_{15}| = 3+\alpha$. Here, $\alpha$ is the penalty coefficient of transfer\footnote{A common practice in transportation research is to set this value to 2.}. 

In addition to the length of a route, we also denote $T_{r_{ij}}$ as the corresponding travel time required when a commuter takes route $r_{ij}$ to travel from the boarding station $s_i$ to the alighting station $s_j$. When the context is clear, we can use $T_{r}$ to represent $T_{r_{ij}}$ for brevity. A trip starts when a commuter enters the turnstile, which consists of following four components, walking to the platform, waiting for the train, travelling via the train, and walking to the turnstile to exit the station and complete the trip. If the route taken requires transfers, an additional component (i.e., transfer) is involved. Accordingly, we can model $T_r$ based on following five kinds of \emph{travel links}, representing the five different travel components described above. In the rest of the paper, the term \emph{travel link} is used to refer to one component of a trip via a metro system, which contributes to the total time required by a trip from entering the boarding station to exiting the alighting station. 
\begin{itemize}
  \item $T_{s_i}^{g}$: entry walking time from turnstiles to the platform at the boarding station $s_i$
  \item $T_{l_x}^w$: waiting time for the train service $l_x$ at station $s_i$
  \item $T_{e}^{c}$: train travel time of every edge $e$
  \item $T_{s}^{q}$: transfer time required at an interchange station $s$
  \item $T_{s_j}^a$: exit walking time from the platform to turnstiles at the alighting station $s_j$
\end{itemize}

\begin{figure*}[t!]
\centering
\includegraphics[width=14cm]{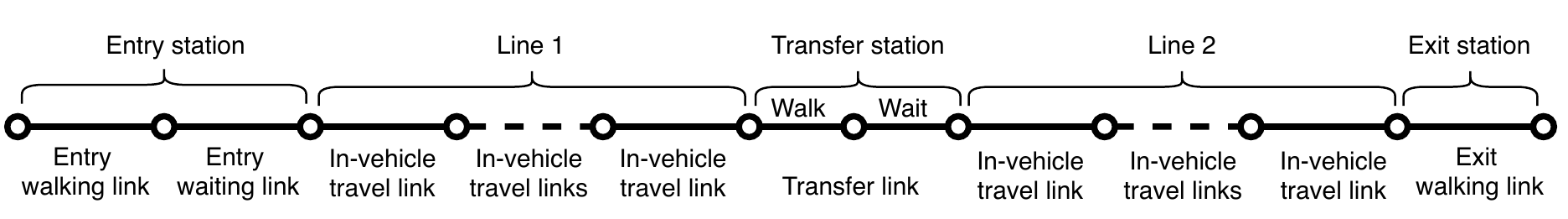} 
\caption{Travel links of a trip in the metro system}
\label{fig:illustration_path} 
\end{figure*}

Here, the transfer time at station $s$ consists of walking time from one platform to another, and waiting time for the next train. For the case of Singapore, most interchanges are crossed by two different metro lines. The only exception is the Dhoby Ghaut station in city center that is crossed by three MRT lines, thus it has three unique transfer walking time distributions. In addition, $T_{l_x}^w$ is independent of the station, as the service frequency of a service line $l_x$ does not change from station to station. Notice in reality, the entry/exit walking time at station $s$ also depends on the platform and turnstiles the commuters travel between, while we abuse the notation here for simplicity. Given an edge $e(s_i,s_j,l_x)$ connecting station $s_i$ and station $s_j$, we assume the travel time required from $s_i$ to $s_j$ via service line $l_x$ is exactly the same as that required from $s_j$ to $s_i$ via the same edge. In other words, we assume that bi-directional travel costs between two adjacent stations are characterized by an identical distribution. However, \solution\ could be easily extended to perform the inferences when we model a metro system as a directed graph, and the travel time required from station $s_i$ to its adjacent station $s_j$ might be different from that from $s_j$ to $s_i$. We further visualize the travel links in Figure~\ref{fig:illustration_path} to facilitate the understanding.   

After decomposing a trip into five different types of travel links, 
we can sum the time spent on each travel link of $r_{ij}$ in order to calculate the total travel time of $r_{ij}=\langle e_{1},e_{2},\cdots,e_{k}\rangle$, as shown in Equation~(\ref{equ:Tr}). Note, $S_m$ refers to the set of interchange stations on route $r_{ij}$ where commuters have to make transfers. 
\begin{equation}T_r = T_{s_i}^{g}+T_{l_x}^w+\sum_{b=1}^{k}{T_{e_b}^{c}}+\sum_{s\in S_m}{T_s^{q}}+T_{s_j}^a
\label{equ:Tr}
\end{equation}

The AFC system records the trips of individual commuters. Each trip observation, represented as $(id,s_o,s_d,t_o,t_d)$, captures the details of a real trip $tr$ made by a commuter via the metro network. Here, $id$ is an encrypted unique string identifying a smart card, $s_o$ is the origin station, $s_d$ is the destination station, $t_o$ records the time stamp when the commuter enters the station $s_o$, and  $t_d$ records the time stamp when the commuter exits the alighting station $s_d$ from a turnstile. In other word, $t=t_d - t_o$ captures the real travel time required. In the rest of this paper, we represent each trip record as $tr=(id,s_o,s_d,t)$ for convenience. Given an OD pair $\langle o,d\rangle$, we collect all the trip records $tr$ that are corresponding to $\langle o, d\rangle$ into a set $TR_{od}$, i.e., $\forall tr \in TR_{od}$, $tr.s_o = o \land tr.s_d=d$. 

This paper aims at inferring the time corresponding to each travel link in order to infer the time required by all possible routes, as well as the probabilities that commuters choose each candidate route to travel for any given OD pair $\langle o, d\rangle$, given a static MRT network structure $G(S, E, L)$ and the set of trip observations captured by the AFC system. We formally define the first inference task in Definition~\ref{defn:infer-time}; we will present the formal definition of the second inference task in Section~\ref{subsec:candidate_route}, after we introduce the concept of candidate route set. 

\begin{defn}\label{defn:infer-time} \textbf{Inference of Route Travel Time.} Given a metro network $G(S, E, L)$, and a large set of trips corresponding to different OD pairs captured by AFC systems $X = \bigcup_{\langle o,d\rangle \in S\times S \land o\ne d} TR_{od}$,
%
%
inference of route travel time is to infer the travel time of all the travel links that might contribute to the total travel time required by any route $r_{ij}$, which can best fit the traveling time observed in $X$. $\square$
\end{defn}

\subsection{Candidate Routes Extraction}
\label{subsec:candidate_route}

As mentioned previously, redundant design is adopted to tolerate faults, because of the reliability requirement of a metro network. Consequently, 
there could be multiple routes available from an origin station $s_i$ to a destination station $s_j$. We therefore introduce the concept of \emph{candidate route set}. Let $R_{od}$ denote the complete set of possible routes of an OD pair $\langle o, d\rangle$, and let $r_{od}^{min}$ refer to the one with the shortest length, i.e., $\forall r \in R_{od}$, $|r_{od}^{min}| \le |r|$ $\land$ $\exists r' \in R_{od}$, $r' = r_{od}^{min}$. Formally, we name $R_{od}$ as the \emph{candidate route set} corresponding to the OD pair $\langle o, d\rangle$. 

To generate a candidate route set $R_{od}$ for each OD pair $\langle o, d\rangle$, there are different strategies, such as edge elimination and $k$-shortest-paths. Nevertheless, the number of stations in a metro system usually is in the scale of either tens or hundreds, e.g., New York City Subway has in total 400+ stations, the most stations owned by a metro system. Consequently, we can simply adopt brute-force-search algorithm to form $R_{od}$ for different OD pairs $\langle o, d\rangle$s. 

Given a candidate route set $R_{od}$ w.r.t. an OD pair $\langle o, d\rangle$, we also notice that some routes may never be used by commuters, e.g., those that are much longer than other routes, and those with too many transfers that bring inconvenience. We, therefore, define a \emph{restricted candidate route set} $R_{od}^{'}$ w.r.t. an OD pair $\langle o, d\rangle$, which excludes those rarely-used or never-used routes based on following criteria:
\begin{itemize}
    \item routes with any loops
    \item routes that are more than $\beta$ ($>1$) times longer than the shortest route $r_{od}^{min}$
    \item routes that are not the shortest paths but require more than $\sigma$ transfers
\end{itemize}

The controlling parameters $\beta$ and $\sigma$ could be set according to different assumptions made on commuters' behavior. For example, in our study, we set both $\beta$ and $\sigma$ to two. The underlying assumptions are i) commuters might not always take the shortest route, but they are not willing to take routes that are much longer than necessary; and ii) some commuters may be willing to make transfers for comfortability or other reasons, but a route that requires more than two transfers is not preferred. However, the solution proposed in this paper is independent on the values of $\beta$ or $\sigma$. 

In brief, $R_{od}^{'}=\{r \in R_{od}| (r = r_{od}^{min} \lor r.q \leq \sigma) \land |r| \leq \beta \times r_{od}^{min} \}$. In the rest of this paper, we refer candidate routes set of an OD pair $\langle o, d\rangle$ to its restricted candidate route set $R_{od}^{'}$. The notation $|R_{od}|$ stands for the number of routes inside the candidate route set $R_{od}$. Based on the candidate route set, we present the second inference task that this paper wants to perform in Definition~\ref{defn:preference}. Note that we can infer the route preference at either the aggregation level or the individual level. At the aggregation level, we can learn the route preferences of the entire commuter population based on the city-scale trip observations; at the individual level, we can learn the preference of a particular individual based on the trip records corresponding to that individual only. 

\begin{defn}\label{defn:preference} \textbf{Inference of Route Preferences.} Given a metro network $G(S, E, L)$ and a large set of trips corresponding to different OD pairs captured by AFC systems 
$X = \bigcup_{\langle o,d\rangle \in S\times S \land o\ne d} TR_{od}$,
%
inference of route preferences is to infer, for an OD pair $\langle o,d\rangle$ with multiple routes (i.e., $|R_{od}|>1$), the likelihood that each route $r\in R_{od}$ will be taken by a commuter to travel from $o$ to $d$. $\square$
\end{defn}

\begin{figure}[t!]
\centering
\includegraphics[width=9cm]{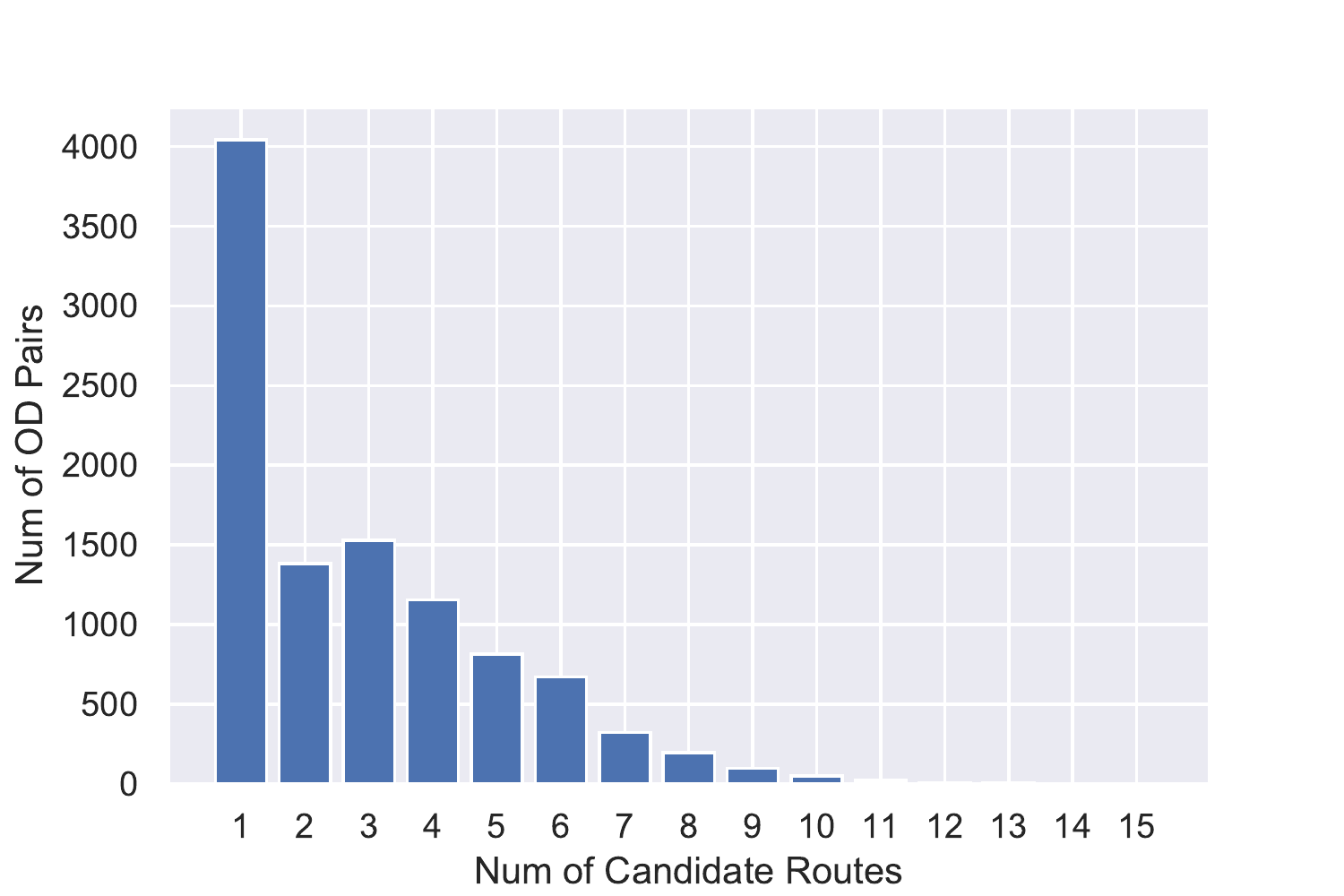}
\caption{Candidate routes distribution of Singapore MRT ($\beta=2 \land \sigma=2$)}
\label{fig:num_of_routes}
\end{figure}

For illustration purpose, we report the size of candidate route sets of different OD pairs corresponding to Singapore metro system in Figure~\ref{fig:num_of_routes}. As it can be observed, the number of candidate routes of OD pairs varies from 1 to 15. We then categorize all the OD pairs according to the sizes of their respective candidate route sets. To be more specific, \emph{single route set} $OD_{s}$ keeps all the OD pairs with single route, and \emph{multiple route set} $OD_{m}$ keeps all the OD pairs with multiple routes, i.e., $OD_s = \{\langle o,d\rangle \in S\times S|o\ne d \land |R_{od}|=1\}$, and $OD_m = \{\langle o,d\rangle \in S\times S|o\ne d \land |R_{od}|> 1\}$. The inference of route preference only focuses on OD pairs preserved by $OD_m$. 

\subsection{Data Exploration Insights}
\label{subsec:insight}

As we highlight in Section~\ref{sec:introduction}, \solution\ adopts an approach that is very different from all the existing solutions, i.e., decoupling the inference of the route travel time from the inferring of the route preferences. To the best of our knowledge, this is the first work that decouples these two inference tasks, which in turn benefits both the accuracy and the efficiency of the inferences. It is worth noting that although decoupling sounds simple, it is non-trivial to propose a two-step framework to not only simplify the inference tasks but also improve the accuracy, as these two inference tasks have interrelationship. Our design is partially motivated by the insights we have collected from Singapore AFC data in our data exploration, to be detailed next. 

\begin{figure*}[t!]
\centering
\includegraphics[width=13cm]{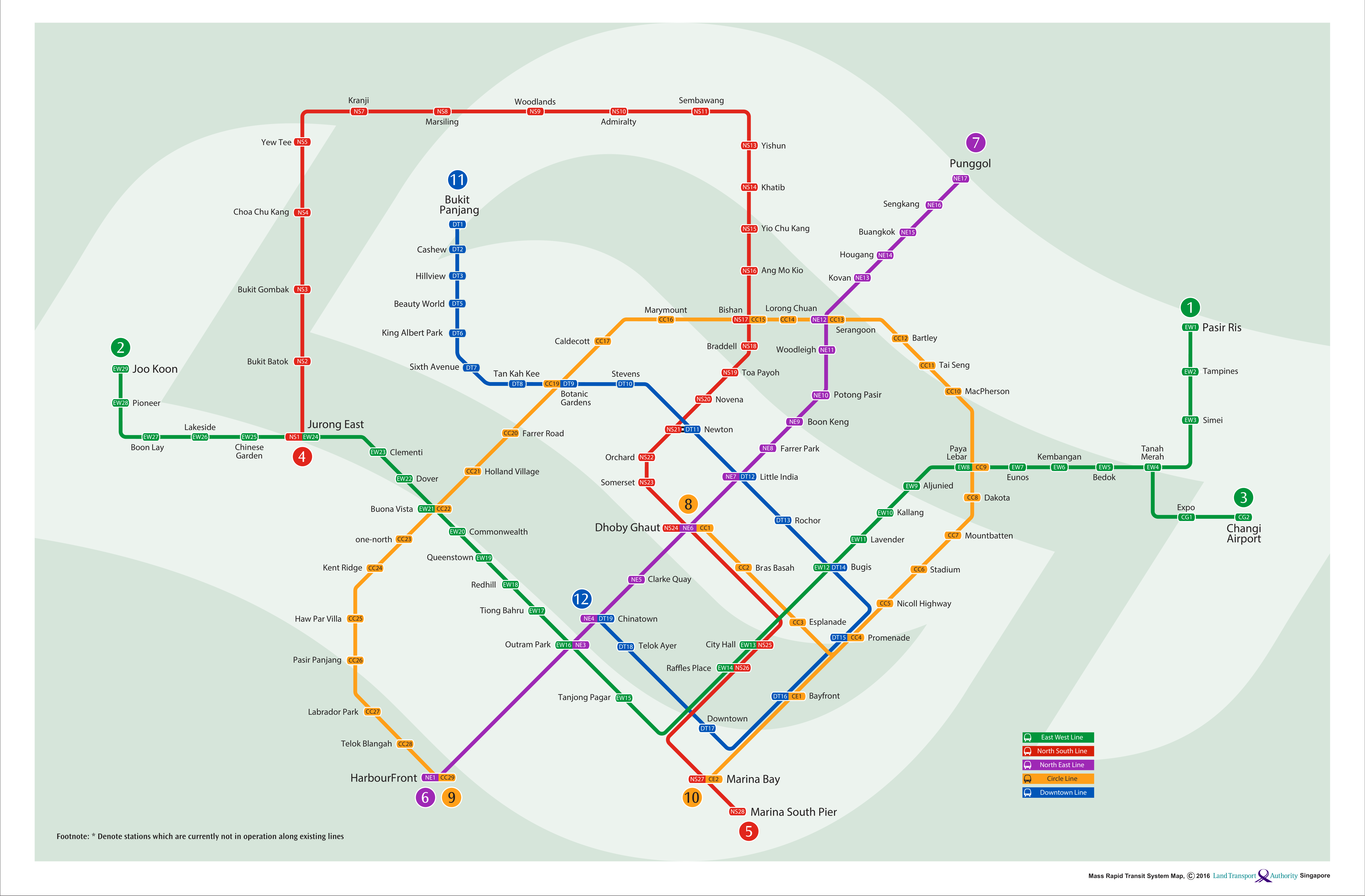}
\caption{Singapore MRT network map (as of 2016 May)}
\label{fig:mrt_map}
\end{figure*}

The Singapore MRT network, as shown in Figure~\ref{fig:mrt_map}, consists of $102$ stations, $7$ MRT lines (including two line extensions), and $114$ edges between adjacent stations (until May 2016). Trip data collected by the AFC system in 2015 December is utilized as one data source in our study. As train operation timetable differs from peak hours to non-peak hours and from weekday to weekend, we study trips happened during weekday morning peak and evening peak respectively, i.e., in Singapore, morning peak is from 7:30am to 9:30am, and evening peak is from 5:30pm to 7:30pm. 

\begin{table}[t!]
\centering
\normalsize
\caption{Statistics of Candidate Routes Sets for Singapore and Taipei}
\label{tab:stats_route_set}
\begin{tabular}{|l|l|l|l|l|}
\hline
&\multicolumn{2}{c|}{Singapore}&\multicolumn{2}{c|}{Taipei}\\
\cline{2-5}
 & $OD_s$ & $OD_m$ & $OD_s$ & $OD_m$\\
\hline
Number of OD pairs & 4,042 & 6,260 & 7,258 &4,298\\ \hline
\% of OD pairs & 39.24\% & 60.76\% & 62.81\% & 37.19\%\\ \hline
\end{tabular}
\end{table}

\begin{table}[t!]
\centering
\normalsize
\caption{Number of Trips Covering Individual Travel Links}
\label{tab:travel-links}
\begin{tabular}{|l|l|l|}
\hline
&\multicolumn{2}{l|}{Number of travel links}\\
\cline{2-3}
Number of trips & Singapore & Taipei \\
\hline
(0,100] & 10 & 6 \\ \hline
(100,1K] & 21 & 34\\ \hline
(1K, 10K] & 93 & 166 \\ \hline
(10K, 100K] & 197 & 205 \\ \hline
>100K & 127 & 82\\ \hline
\end{tabular}
\end{table}

There are in total $10,302$ (i.e., $=101\times 102$) OD pairs inside the Singapore MRT network. After generating candidate route sets for all OD pairs as described in Section~\ref{subsec:candidate_route}, we find that $39.24\%$ of OD pairs have only one candidate route, i.e., single route set $OD_s$ consists of $10,302\times 39.24\% = 4,042$ OD pairs, as reported in Table~\ref{tab:stats_route_set}. When we further check those $4,042$ OD pairs in $OD_s$, we find out that their routes actually cover each single travel link that might be a component of the travel time $T_r$ of any route $r$ (i.e., a component of Equation~(\ref{equ:Tr})). 

To be more specific, given a metro system, we could enumerate all the travel links. Take Singapore MRT network as an example. There are $7$ service lines, so there are in total $7$ travel links corresponding to $T^w_{l_x}$. There are $114$ edges, so there are in total $114$ travel links corresponding to $T^c_e$. There are $102$ stations with $19$ being interchanges and $83$ being normal stations. Each normal station contributes one travel link to $T^g_{s_i}$ and one travel link to $T^a_{s_i}$, while each interchange could produce multiple travel links to $T^g_{s_i}$ and $T^a_{s_i}$, dependent on the number of the platforms and the number of exits it has. Take Dhoby Ghaut station as an example. It is passed by $3$ lines and has $2$ exits located at very different locations, where each unique platform-exit combination produces a unique travel link, so in total it contributes to $3\times 2=6$ travel links to both $T^g_{s_i}$ and $T^a_{s_i}$. In summary, there are $153$ travel links corresponding to $T^g_s$ and $T^a_s$ respectively. The number of travel links corresponding to $T^q_s$ depends on the number of lines passing by each interchange station, and the number of interchange stations, and in total there are 21 travel links. 
%
%
In other words,  we have $(7+114+153\times 2+21)=448$ travel links corresponding to the Singapore metro system (as of May 2016). If we could derive all those $448$ travel links, the travel time of any route could be recovered based on Equation~(\ref{equ:Tr}). The real trip records corresponding to single route OD pairs captured by the AFC system actually cover each single travel link. Here, we say a trip record covers a travel link if and only if the travel link contributes to the time duration required by the trip. 

This observation suggests a possibility that we actually have sufficient trip observations to perform inference of route travel time based \emph{only} on the trips corresponding to the OD pairs in the single route set $OD_s$. Recall that all the OD pairs $\langle o,d\rangle$s in the single route set $OD_s$ share a common unique feature, that is there is only one route sending a commuter from the origin station $o$ to the destination station $d$. Accordingly, given an AFC trip record $tr$ from $o$ to $d$, we know the exact route taken by the commuter for the trip $tr$. This is to say, we can locate all the travel links travelled by $tr$ without any ambiguity, i.e., the entry/exit station, the links $e_1$, $e_2$, $\cdots$, $e_k$ travelled and the transfer $S_m$ required in Equation~(\ref{equ:Tr}) are known. In other words, we can take all the trip records $tr$s that are corresponding to OD pairs inside the single route set $OD_s$ to infer the travel time of different travel links. This significantly simplifies the inferring of the travel time, which will be further demonstrated by our experimental study to be presented in Section~\ref{sec:case_study}. 

To further verify our conjecture and test if there are sufficient trip records to perform the inference, for each travel link in the metro network $G$, we further count the number of trips corresponding to only the single route OD pairs that cover the travel link, as reported in Table~\ref{tab:travel-links}. Notice that the count reported is based on one month of morning peak trip records collected by AFC system in Singapore. As can be observed, $97.8\%$ of the travel links are covered by more than $100$ trip records, and $93.1\%$ of the travel links are covered by more than $1,000$ trip records, which suggests that the trips corresponding to single-route OD pairs are indeed sufficient to perform robust inferences of the travel time of each travel link. Note that the number of trips will be further increased when the duration corresponding to the data collection is extended. Consequently, we would like to conclude that the finding of the trip records of single-route OD pairs being sufficient to infer the travel time is NOT a coincidence that is only observed from the Singapore dataset. For example, we have performed a similar study on Taipei dataset, again based on one month city-scale data collection. As reported in Table~\ref{tab:stats_route_set} and Table~\ref{tab:travel-links}, the above statement is also valid on Taipei dataset. 

\section{Solution Algorithms}
\label{sec:solution_algorithm}

\begin{figure*}[t!]
\centering
\includegraphics[width=10cm]{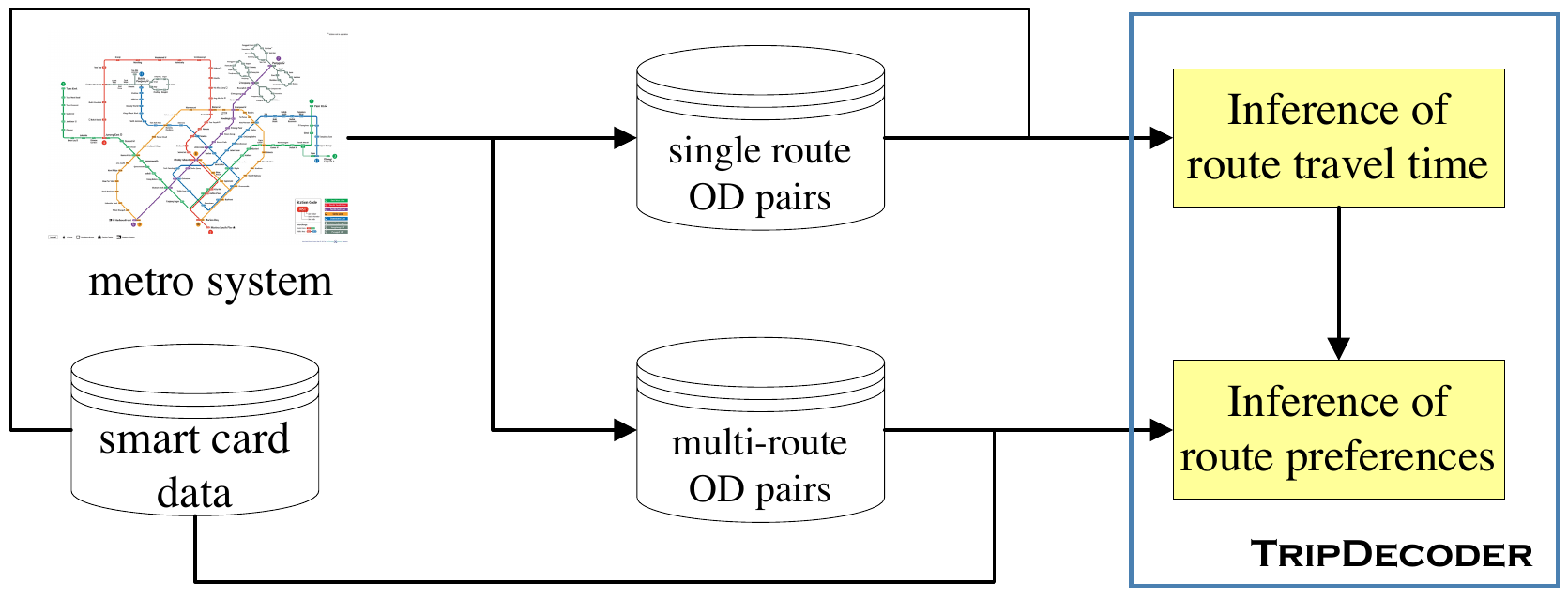}
\caption{\solution\ framework}
\label{fig:model_framework}
\end{figure*}

As highlighted before, we propose to decouple the inference of route travel time from the inference of route preferences. Accordingly, there are two major components in \solution, the framework proposed in this paper to perform the inference tasks. Figure~\ref{fig:model_framework} depicts the architecture of \solution. In the following, we detail how \solution\ performs these two inference tasks. 

\subsection{Travel Time Inference}
\label{subsec:travel_time_inference}

As stated in Equation~(\ref{equ:Tr}), the travel time of a trip consists of five types of travel links, represented by $T_s^g$, $T_l^w$, $T_e^c$, $T_s^q$ and $T_s^a$ respectively. In this project, we assume $T_s^g$,  $T_l^w$, $T_e^c$, $T_s^q$ and $T_s^a$ all follow normal distribution, due to its simplicity and additive properties.
\begin{align}
T_s^g& \sim N(\mu_s^g,\sigma_s^g)\\ 
T_l^w &\sim N(\mu_l^w,\sigma_l^w) \\
T_e^c &\sim N(\mu_e^c,\sigma_e^c) \\
T_s^q &\sim N(\mu_s^q,\sigma_s^q) \\
T_s^a& \sim N(\mu_s^a,\sigma_s^a)
\end{align}

where the probability distribution function of Gaussian distribution $N(\mu,\sigma)$ is defined as 
\begin{align}
N(x;\mu,\sigma) = \frac{1}{\sqrt{2\pi\sigma^2}}e^{\frac{-(x-u)^2}{2\sigma^2}}
\end{align}

As mentioned before, we assume the waiting time of a service line $l$ is independent on the stations, while different MRT lines can have different $\{\mu_{l}^{w},\sigma_{l}^{w}\}$. 

For the inference of route travel time, it has to learn the mean and the variance in the normal distribution for each travel link.
%
%
The travel time from station $i$ to station $j$ which follows a normal distribution is denoted by $T_{ij} \sim N(\mu_{ij},\sigma^2_{ij})$. Since $T_{ij}$ assembles the travel time of every travel link covering the route, the distribution of $T_{ij}$ can be approximated by a Gaussian distribution and the mean $\mu_{ij}$ and variance $\sigma^2_{ij}$ could be derived as follows: 
\begin{align}
T_{ij}& \sim N(\mu_{ij},\sigma_{ij}) \\
\mu_{ij}& =  \mu_{s_i}^{g}+\mu_{l_x}^w + \sum_{b=1}^{K}\mu_{e_b}^{c} + \sum_{s\in S_m}\mu_s^q + \mu_{s_j}^a\\ 
\sigma_{ij}^2& =  {\sigma_{s_i}^{g}}^2+{\sigma_{l_x}^w}^2 + \sum_{b=1}^{K}{\sigma_{e_b}^c}^2 + \sum_{s\in S_m}{\sigma_s^q}^2 + {\sigma_{s_j}^a}^2 
\end{align}

For OD pair $\langle i, j \rangle$ with a single route (i.e., $\langle i, j \rangle \in OD_s$), the likelihood of observing travel time $t$ is
\begin{align}
L(\mu_{ij},\sigma_{ij}^2;t)=N(t;\mu_{ij},\sigma_{ij}^2)=\frac{1}{\sqrt{2\pi\sigma_{ij}^2}}e^{\frac{-(t-u_{ij})^2}{2\sigma_{ij}^2}}
\end{align}


Let parameter set $\Theta$ include all the travel time parameters to be inferred, i.e., $\{\mu_s^g,\sigma_s^g\}$, $\{\mu_s^q, \sigma_s^q\}$, and  $\{\mu_s^a,\sigma_s^a\}$ for all the stations $s \in S$, $\{\mu_e^c,\sigma_e^c\}$ for all the edges $e \in E$, and $\{\mu_l^w,\sigma_l^w\}$ for all the lines $l \in L$. We propose to use maximum likelihood methods to estimate them. Given a set of history trip records $TR_s$ corresponding to the OD pairs in the single route set $OD_s$, i.e., $TR_s = \{(id_i, o_i, d_i, t_i)|\langle o_i, d_i\rangle \in OD_s\}$, the full likelihood of $TR_s$ is
\begin{align}
\label{eq:s1_likelihood}
L(\Theta;TR_s) = \prod_{i=1}^{|TR_s|}\left [
\frac{1}{\sqrt{2\pi\sigma_{o_id_i}^2}}e^{\frac{-(t_i-u_{o_id_i})^2}{2\sigma_{o_id_i}^2}}
\right ]
\end{align}

Taking the logarithm of Equation~(\ref{eq:s1_likelihood}), we can get the log-likelihood as

\begin{align}
\label{eq:s1_log_likelihood}
l(\Theta;TR_s)
= \sum_{i=1}^{|TR_s|}\left [-\frac{1}{2}\ln(2\pi\sigma^2_{o_id_i})-\frac{(t_i-\mu_{o_id_i})^2}{2\sigma_{o_id_i}^2}\right ]
\end{align}

To begin with, we initialize $\Theta$ based on prior knowledge of the metro network or empirical observations. Take Singapore MRT network as an example. $\mu_e^c$ is set based on statistics provided by Singapore Land Transport Authority (LTA), and the corresponding $\sigma_e^c$ is set to $\mu_e^c/10$. $\{\mu_s^g,\sigma_s^g\}$, $\{\mu_s^q, \sigma_s^q\}$, $\{\mu_s^a,\sigma_s^a\}$ and $\{\mu_l^w,\sigma_l^w\}$ are set based on empirical observations as shown in Table~\ref{tab:default_parameters}. Thereafter, we perform stochastic gradient descent (SGD) to tune the parameters by maximizing $l(\Theta;TR_s)$. 

\begin{table}[t!]
\centering
\normalsize
\caption{Initial Values for Travel Steps}
\label{tab:default_parameters}
\begin{tabular}{|c|c|c|}
\hline
\textbf{Travel link} & \textbf{$\mu$ (seconds)} & \textbf{$\sigma$ (seconds)} \\ \hline
Normal station entry/exit walking & 60 & 12 \\ \hline
Interchange entry/exit walking & 120 & 24 \\ \hline
Transfer walking & 60 & 12 \\ \hline
Train service waiting & \textit{tf} * 0.5 & \textit{tf} * 0.05\\ \hline
\multicolumn{3}{l}{$^{\mathrm{a}}$\textit{tf} represents train frequency of a MRT line}\\
\end{tabular}
\end{table}

\subsection{Route Preferences Inference}
\label{subsec:route_choice_preference_inference}

\begin{figure}[t!]
\centering
\includegraphics[width=10cm]{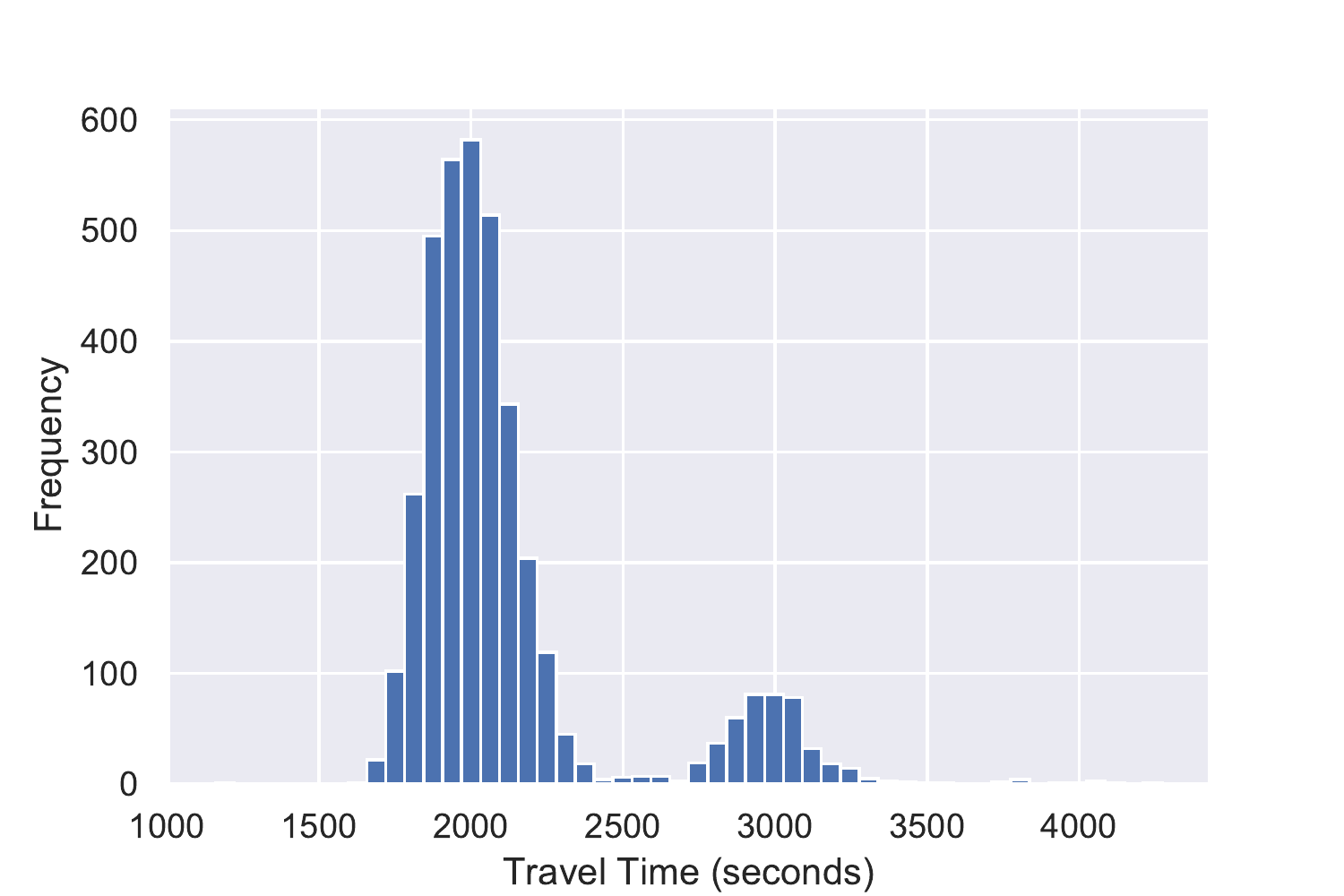}
\caption{Travel time observations from Bishan to Jurong East}
\label{fig:bisan_jurong_stats}
\end{figure}

\begin{figure}
    \centering
    \begin{minipage}{0.45\textwidth}
        \centering
        \includegraphics[width=0.95\textwidth]{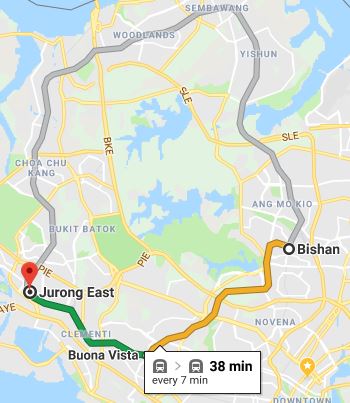} 
        Route 1
    \end{minipage}
    \begin{minipage}{0.45\textwidth}
        \centering
        \includegraphics[width=0.95\textwidth]{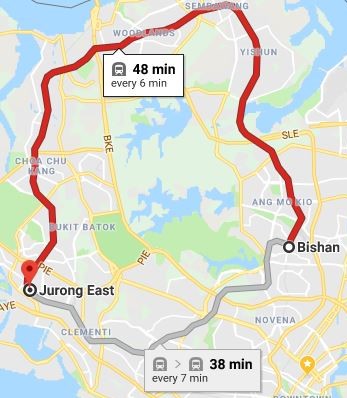} 
        Route 2
    \end{minipage}
    \caption{Two candidate routes from Bishan to Jurong East on Google Map}
    \label{fig:bishan_jurong_google}
\end{figure}

Based on travel time deduced in Section~\ref{subsec:travel_time_inference}, we are ready to assemble travel time distribution of any route. For a given OD pair $\langle o,d \rangle \in OD_m$, each candidate route in $R_{od}$ has a unique travel time distribution. Take the routes from Bishan station to Jurong East station as an example. The real travel time distribution based on $TR_{od}$ is depicted in Figure~\ref{fig:bisan_jurong_stats}, where we could observe two different patterns, one pattern having an average travel time of about $33$ minutes, and the other having an average travel time of about $50$ minutes. It is very likely that these two patterns of the travel time represent the two different candidate routes, as suggested by Google Map shown in Figure~\ref{fig:bishan_jurong_google}. Thus, $TR_{od}$ can be modeled as a mixture of distributions from the different candidate routes,
\begin{align}
TR_{od} &\sim \sum_{r\in R_{od}}\pi(r)N(t;\mu_r,\sigma_r^2)
\end{align}
where $\mu_r$ and $\sigma_r$ of each candidate route have already been derived in Section~\ref{subsec:travel_time_inference}. $\pi(r)$ refers to the probability that commuters take $r$ when traveling from $o$ to $d$ and thus $\sum_{r\in R_{od}}\pi(r)=1$.

Assume we have a set of historical trip records $TR_m$ corresponding to the OD pairs in the multiple route set $OD_m$, i.e., $TR_m = \{(id, o_i, d_i, t_i)|$ $\langle o_i, d_i\rangle \in OD_m\}$. Let $\Pi$ represent the set of parameters related to route preference to be derived, i.e., $\Pi = \{pi(r)|r \in R_{od} \land |R_{od}|>1\}$. Then, the full likelihood of $TR_m$ can be written as,

\begin{align}
\label{eq:s2_likelihood}
L(\Pi;TR_m) = \prod_{i=1}^{|TR_m|}[\sum_{r\in R_{o_id_i}}\pi(r)N(t_i;\mu_r,\sigma_r^2)]
\end{align}

Taking the logarithm of Equation~(\ref{eq:s2_likelihood}), we get the log-likelihood as 
\begin{align}
\label{eq:s2_log_likelihood}
l(\Pi;TR_m) = \sum_{i=1}^{|TR_m|}\ln(\sum_{r\in R_{o_id_i}}\pi(r)N(t_i;\mu_r,\sigma_r^2))
\end{align}

To begin with, we assign equal likelihood to every candidate route of any OD pairs in $OD_m$, i.e.,
$$
\pi(r) = 1/|R_{od}|, \forall r \in R_{od}\ \land\ \forall \langle o,d \rangle \in OD_m 
$$

To get the maximum of Equation~(\ref{eq:s2_log_likelihood}), again we perform SGD to tune route preferences $\Pi$.

\section{Case Study}
\label{sec:case_study}

To demonstrate the superior performance of \solution, we conduct studies based on two real sets of city-scale trip records collected in Singapore and Taipei. As mentioned before, we consider both accuracy and efficiency as the goals when we design \solution. Consequently, we will compare the accuracy and the efficiency of \solution\ with the state-of-the-art works. In the following, we first briefly explain the real datasets used in this study, and then present the two sets of experiments that evaluate the accuracy and the efficiency of \solution\ as well as other state-of-the-art works. 
%

\subsection{Experiments Settings}
\subsubsection{Data Brief}

\renewcommand{\tabcolsep}{1pt}
\begin{table}[t!]
\centering
\caption{Smart Card Data Sample}
\label{tab:ezlink_sample}
\begin{tabular}{|p{1.2cm}|p{1cm}|p{1.8cm}|p{1.8cm}|p{1.7cm}|p{1.7cm}|p{1.5cm}|p{2.3cm}|}
\hline
\textbf{id} & \textbf{type} & \textbf{entry date} & \textbf{entry time} & \textbf{exit date}  & \textbf{exit time}  & \textbf{origin id} &\textbf{destination id}                                                   
\\ \hline
02***5F & adult & 2015-12-02 & 08:20:04 & 2015-12-02 & 08:27:27 & 35 & 12
\\ \hline
02***5F & adult & 2015-12-02 & 18:13:57 & 2015-12-02 & 18:21:25 & 12 & 35
\\ \hline
02***5F & adult & 2015-12-03 & 08:13:51 & 2015-12-03 & 08:21:21 & 35 & 12
\\ \hline
02***5F & adult & 2015-12-03 & 18:31:45 & 2015-12-03 & 18:38:11 & 12 & 35
\\ \hline
02***5F & adult & 2015-12-03 & 18:47:45 & 2015-12-03 & 19:01:16 & 35 & 12
\\ \hline
\end{tabular}
\vspace{0.1cm}
\end{table}

\begin{table}[t!]
\centering
\normalsize
\caption{Smart Card Dataset Attributes}
\label{tab:ezlink}
\begin{tabular}{|p{2.1cm}|p{1.6cm}|p{8cm}|}
\hline
\textbf{attribute}           & \textbf{notation} & \textbf{description}                                                                                \\ \hline
id           & $c_{id}$        & unique identifier of a smart card                                                                    \\ \hline
type & $type$ & commuter type (i.e., child, adult, senior) \\  \hline
entry date               & $date_{in}$          & starting date of a ride                                                                             \\ \hline
exit date                & $date_{out}$          & ending date of a ride                                                                         \\ \hline
entry time               & $t_{in}$          & starting time of a ride  \\ \hline
exit time                & $t_{out}$         & ending time of a ride \\ \hline
origin id      & $id_{in}$        & unique identifier of the origin MRT station/bus stop    \\ \hline
destination id & $id_{out}$        & unique identifier of the destination MRT station/bus stop         \\ \hline
\end{tabular}
\end{table}

Our study is based on two sets of city-scale real trip records, captured by AFC systems in Singapore and Taipei respectively. In particular, EZ-Link card and EasyCard are used as the smart cards for payment of public transport in Singapore and Taipei, respectively. EZ-Link card data collected in Singapore in 2015 December, and EasyCard data collected in Taipei in 2018 August are used in this study. As our study is based on metro networks, we exclude the data corresponding to bus rides in this study. 

%
%
As listed in Table~\ref{tab:ezlink_sample}, each EZ-Link record in our data collection is corresponding to one MRT ride, including the boarding and alighting MRT stations and the corresponding timestamps. Other information such as the commuter type and the fare charge are also recorded. Apart from that, each smart card is associated with an encrypted unique identifier, so that we can identify all the rides taken by one commuter with commuter's real identity being well protected. Table~\ref{tab:ezlink} lists the attributes captured by each EZ-Link record. Taipei EasyCard data is no different except that it doesn't release the encrypted unique identifier of each card so we are not able to differentiate the trips made by one commuter from those made by other commuters.

Due to defects of an AFC system, sometimes it generates duplicate records or trip records with unrealistic travel duration (e.g., trips that last more than multiple hours or less than two minutes). These records may bias our analytics and hence are removed. 

\subsubsection{Baselines}

Our proposed method \solution\ is compared against both commercial Apps and academic research works. For commercial Apps, we include \emph{Google Map}, the most popular direction service in both Singapore and Taipei as the main representative. In addition, we also include \emph{Gothere} (https://gothere.sg), a very popular direction service used in Singapore. For academic research works, the inference model published in NIPS 2017~\cite{NIPS2017} is the latest work which is employed as the representative of the state-of-the-art work, denoted as \emph{NIPS}. 

\subsubsection{Metrics.}

We employ the \emph{prediction error} and the \emph{execution time} to evaluate model effectiveness and efficiency respectively. Given a set of trip observations $X$ and a prediction method $\rho$, let $\mu_{od}$ be the expected travel time of route $r_{od}$ predicted by $\rho$. The prediction error of travel time of $\rho$ is defined as:
\begin{equation}\label{equ:error}
error_{X,\rho} = \frac{1}{|X|}\sum_{(s_o,s_d,t)\in X}\frac{|\mu_{od}-t|}{t}
\end{equation}

The execution time reported in this paper is obtained by running the two inference tasks on Microsoft Windows 10 Education instances, each of which is shipped with a Intel Core i8-8700 CPU @3.20GHz and a 32.0GB RAM.

\subsection{Accuracy Evaluation}
\label{subsec:experiment_results}

\solution\ deduces both travel time parameters $\Theta$ and route preference $\Pi$. To report its performance in a more comprehensive way, we conduct different sets of experiments to compare the performance of \solution\ with its competitors, for both the prediction error of derived $\Theta$ (corresponding to the inference of travel time) and that of $\Pi$ (corresponding to the inference of route preferences). 

\begin{table}[t!]
\caption{Travel Time Prediction Error}
\centering
\label{tab:time-prediction-error}
\begin{tabular}{|l||l|l||l|l|}
\hline
 &\multicolumn{2}{c|}{Singapore}&\multicolumn{2}{c|}{Taipei}\\
\cline{2-5}
 & Morning Peak & Evening Peak & Morning Peak & Evening Peak\\
\hline\hline
\solution & 8.53\% & 10.38\% & 10.26\% & 10.88\%\\
NIPS & 17.40\% & 20.19\% & 19.63\% & 21.67\%\\
Google Map & 19.94\% & 19.49\% & 23.67\% & 25.45\%\\
Gothere & 12.06\% & 13.57\% & - & -\\
\hline
\end{tabular}
\end{table}

\vspace{0.1in}
\noindent
\textbf{Evaluation of Travel Time Parameters}. In Section~\ref{subsec:travel_time_inference}, we use trip observations set $TR_s$ to infer travel link time parameters, which are then used to construct travel time of any route. As there is only one route for any OD pair in $OD_s$, there is no ambiguity among the candidate routes so we know exactly the route taken by the commuters to travel from an origin $o$ to a destination $d$ for each $\langle o,d\rangle \in OD_s$. Consequently, we can derive $\mu_{od}$ for each trip record in $TR_s$ and derive the prediction error following Equation~(\ref{equ:error}). 

Table~\ref{tab:time-prediction-error} reports the prediction error of \solution\ and its competitors. For both Google Map and Gothere, we submit 20 queries for each OD $\langle o, d \rangle$ pair in $OD_s$ and report the average performance. Each of those 20 queries has the boarding station $o$ as its current location, alighting station $d$ as its destination, and the trip start time is randomly selected from the duration (e.g., for morning peak in Singapore, the trip start time is randomly selected from 7:30am to 9:30am). As could be observed from the results, \solution\ demonstrates a superior accuracy performance. For example, \solution\ reduces the prediction error of Google Map by 14.25\% and 17.57\%, for Singapore dataset and Taipei dataset respectively. 

\begin{table}[t!]
\caption{Route Preferences Prediction Error (dataset: Singapore)}
\centering
\label{tab:route-preference-error}
\begin{tabular}{|l||l|l|}
\hline
 & Morning Peak & Evening Peak \\
\hline\hline
\solution & 6.20\% & 7.02\% \\
Shortest Route & 7.40\% & 8.00\% \\
NIPS & 15.99\% & 16.52\% \\
\hline
\end{tabular}
\end{table}

\vspace{0.1in}
\noindent
\textbf{Evaluation of Route Preferences}. As stated in Definition~\ref{defn:preference}, the inference of route preferences is to infer the likelihood that a commuter takes a specific route $r \in R_{od}$ when the route candidate set $R_{od}$ has multiple routes. However, we do not have the ground truth of which commuter takes which route to complete her trip inside the metro network. Consequently, how to evaluate the accuracy of the second inference task remains challenging. In this study, we propose a novel evaluation plan that is based on an assumption that a commuter who has to travel from station $s_o$ to station $s_d$ regularly has her own preferences, if there are multiple routes available. The main idea is to learn individual commuter's preference for the trip from station $s_o$ to the station $s_d$ from the historical trips made by the commuter, and to predict the route taken by the commuter when she is about to make the same trip. Based on the route preference, we can infer the time required and get the accuracy when we compare the predicted time with the ground truth time captured by AFC data. 

In order to implement this evaluation plan, we strategically group the trip records in $TR_m$ based on $\langle id, o, d\rangle$. That is to say trip records in the same group, denoted as $TR^{id}_{od}$, share the same $id$, $o$ and $d$ values and we order the groups in descending order of their set sizes $|TR^{id}_{od}|$, the number of trip records in each group. We then select the top 10\% of the groups based on their set size. In Singapore dataset, top $10\%$ groups have their set size ranging between $20$ and $23$. For each of such selected groups $TR^{id}_{od}$, we partition the trip records into two disjoint subsets based on the ratio of $1:1$, with one as the training subset and the other as the testing subset. We then feed the training subset to \solution\ and NIPS for the inference of route preference of the commuter whose encrypted identifier is $id$, and use the inferred route preference to predict the route taken by the commuter for the trips in the testing subset. For each trip $\langle id, o, d\rangle$ in the testing subset, we can predict the time $t_p$ required by this trip, which is derived based on the preferred route $r_{od}$ inferred, the time parameter of travel links inferred in the previous inference of route travel time, and Equation~(\ref{equ:Tr}). Accordingly, given the testing subset, we can measure the prediction error of \solution\ and NIPS, respectively based on Equation~(\ref{equ:error}).


Table~\ref{tab:route-preference-error} shows the prediction error of \solution\ and NIPS by using Singapore dataset. We exclude Taipei dataset from this set of experiments as $id$ is not available in Taipei dataset. Note that in addition to NIPS, we also implement \emph{Shortest Route}, which refers to a very common assumption made by many existing works, i.e., commuters tend to take the shortest route when there are multiple routes available. Again, \solution\ incurs the lowest prediction error. It is worth noting that \solution\ outperforms Shortest Route even in the morning peak, when most of the regular commuters are expected to be more sensitive to the travel time required. It also implies that commuters do not always take the shortest route even during weekday morning peak. 
%

\subsection{Efficiency Evaluation}
\label{subsec:efficiency_comparison}

\begin{figure}[t!]
\centering
\includegraphics[width=10cm]{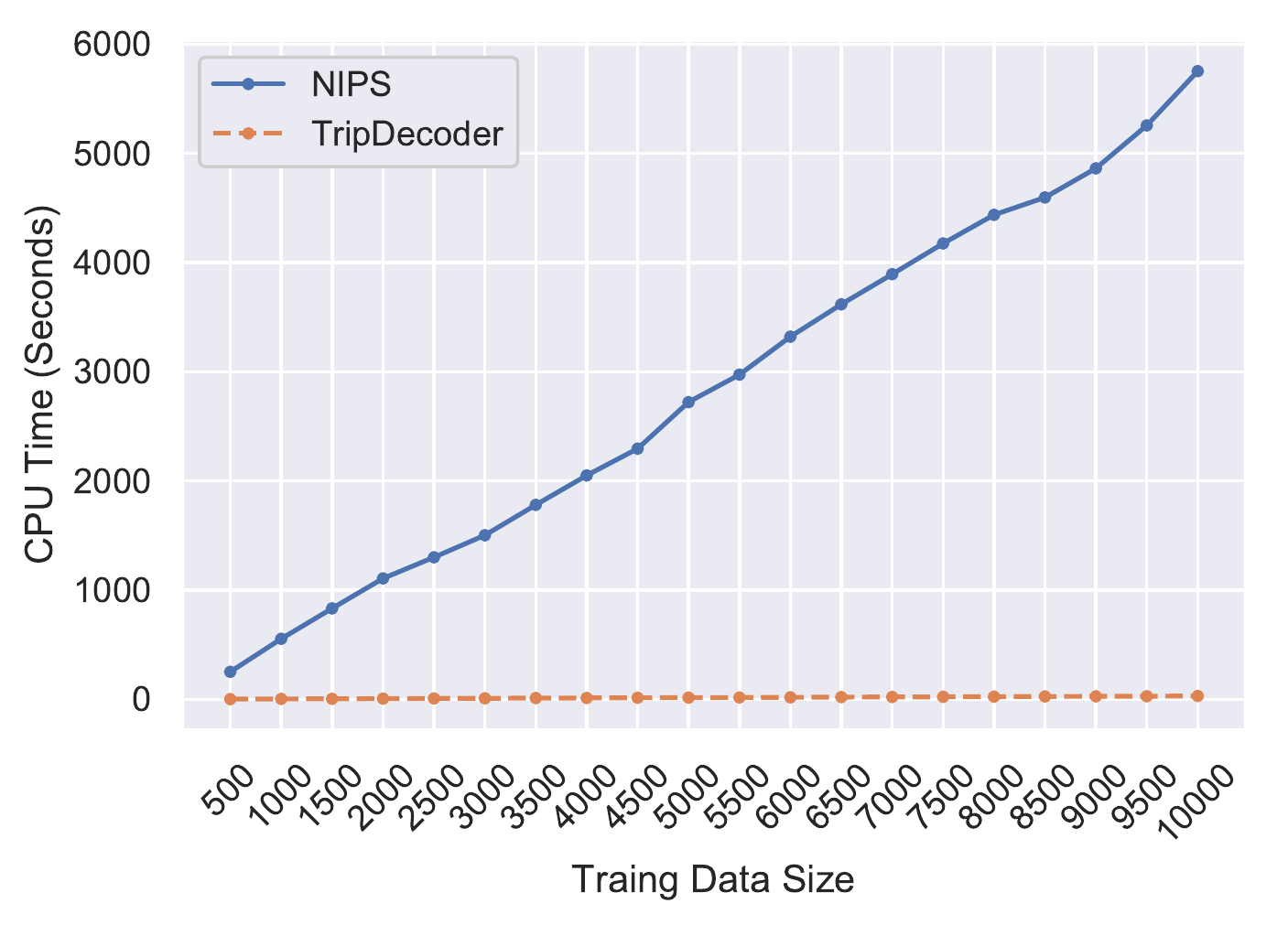}
\caption{CPU time of 50 iterations vs. training data size represented by the total number of trip records}
\label{fig:efficiency_comparison}
\end{figure}


\begin{figure}[htb]%
\centering
    \subfigure[\solution\ vs. NIPS]{
    \label{fig:s1_error-1}
    \psfig{file=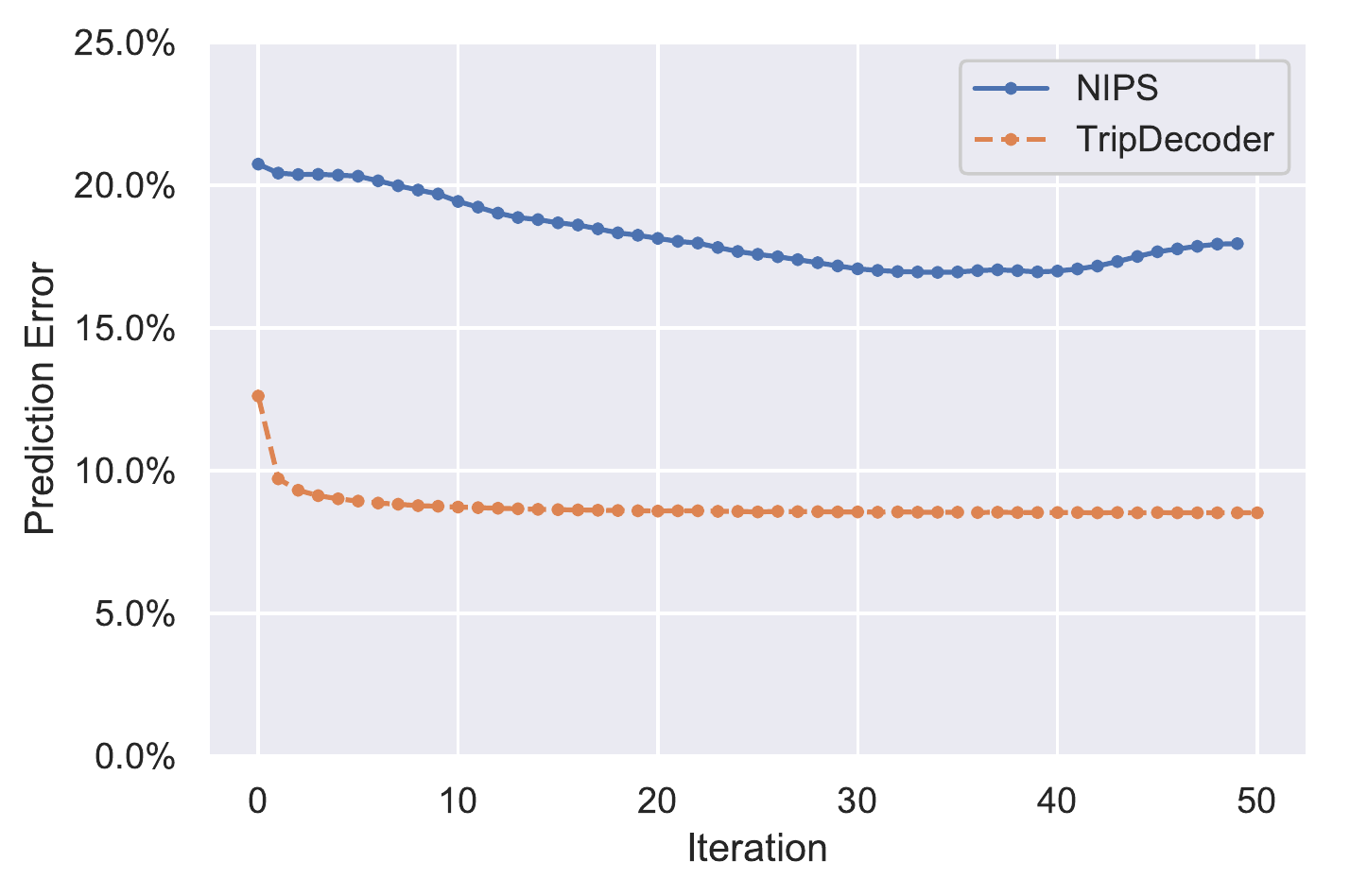, width=0.45\textwidth}
    }
    \centering
    \subfigure[NIPS]{
    \label{fig:s1_error-2}
    \psfig{file=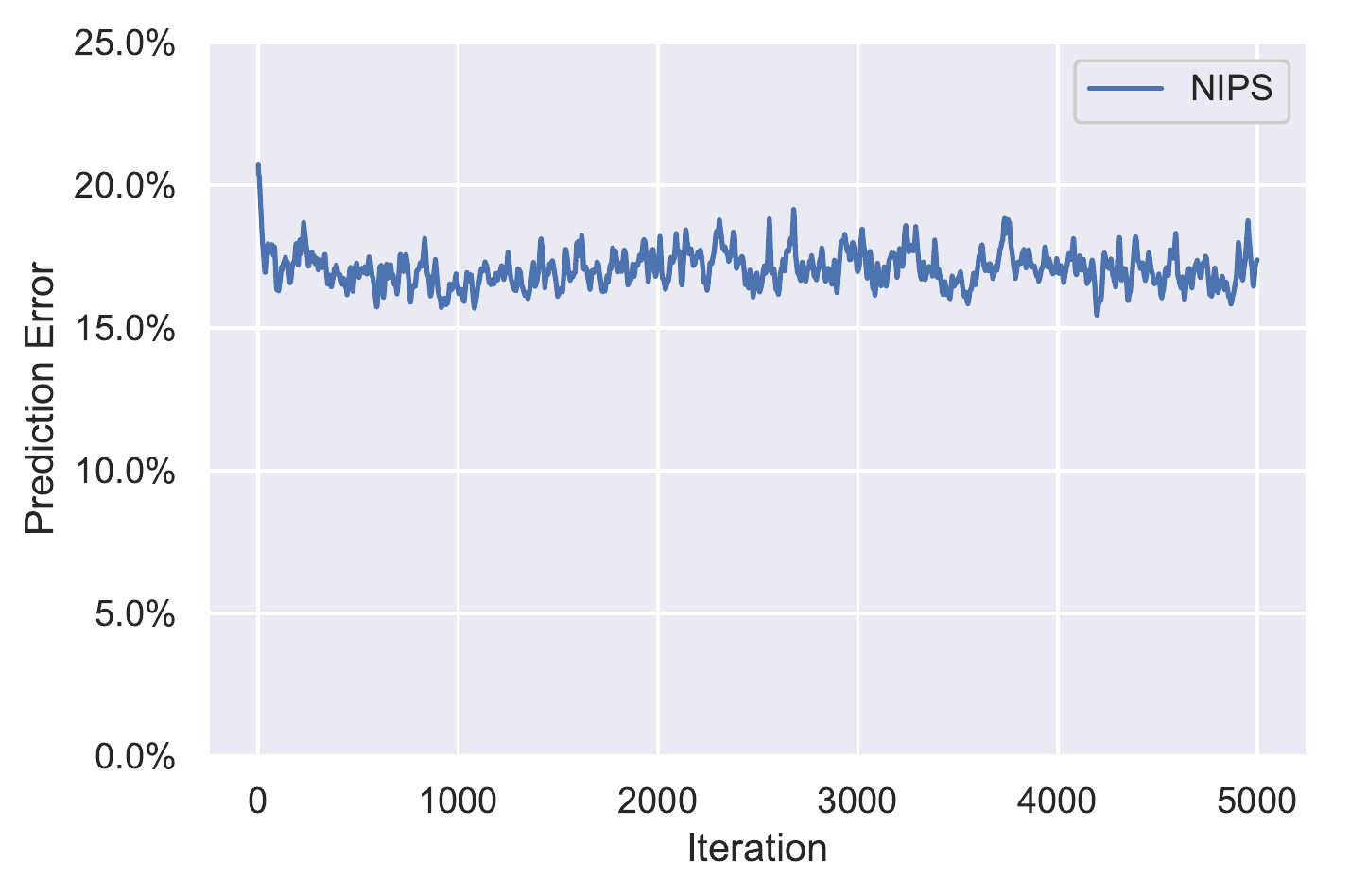, width=0.45\textwidth}
    }
\caption{Travel time prediction error vs. training iterations}
\label{fig:s1_error}
\end{figure}

As we mentioned in Section~\ref{sec:introduction}, \solution\ is designed to not only improve the the accuracy of inference but also to address the efficiency issue. Ideally, we prefer a model that can complete the inference tasks with low prediction error within a short duration of time. Therefore, we can evaluate the efficiency of a model in two perspectives, including \emph{execution time} and the \emph{rate of convergence}. 

Figure~\ref{fig:efficiency_comparison} reports the execution time of 50 iterations with the size of training data varied. It can be seen that the execution time of both models increases linearly.  However, the execution time of \solution\ grows slightly whereas that of NIPS grows significantly. Consequently, \solution\ has a much better scalability and is more suitable to process city-scale trip records. 
%
%

We also study the number of iterations required by the inference models to achieve a stable prediction error and report the result in Figure~\ref{fig:s1_error-1}. First, we could observe that both \solution\ and NIPS are able to achieve higher prediction accuracy as the number of iterations increased. 
%
%
Second, prediction error of \solution\ actually decreases significantly in the first few iterations of training and achieves stable and close-to-optimal performance only after a small number of iterations; whereas NIPS cannot decrease its prediction error in the first 50 iterations steadily. To further investigate the convergence rate of NIPS, we keep tracking the prediction error of NIPS by increasing the number of iterations from $50$ to $5000$, as reported in Figure~\ref{fig:s1_error-2}. It can be observed that the prediction error of NIPS keeps vibrating and tends not to converge. We can conclude that \solution\ demonstrates the superior efficiency and the so-called super convergence capability which is very desirable in model training whereas NIPS results in limited scalabity in practice.


\section{Conclusion}
\label{sec:conclustion}

In this paper, we target at recovering the exact routes taken by commuters inside a metro system that are not captured by an AFC system and hence remain unknown. In 2016, London Tube system run a 4-weeks' trial to log more than 500 million WiFi connection requests from around 5.6 million devices. One of the main objectives was to track the journeys around the network and to recover how commuters move inside the network. Without incurring additional cost, \solution\ is able to achieve the same goal based on available data already captured by an AFC system. 

We strategically propose two inference tasks to handle the recovering, one to infer the travel time of each travel link that contributes to the total duration of any trip inside a metro network and the other to infer the route preference based on historical trip records and the travel time of each travel link inferred in the previous inference task. As these two inference tasks have interrelationship, most of existing works perform these two tasks simultaneously. However, we adopt a totally different approach when we design \solution. \solution, to the best of our knowledge, is the first model that points out and fully utilizes the fact that there are some trips inside a metro system with only one practical route available. It smartly decouples these two inference tasks by only taking those trip records with only one practical route as the input for the first inference task of travel time and feeding the inferred travel time to the second inference task as an additional input which not only improves the accuracy of both inference tasks but also effectively reduces the complexity of the inference tasks. We have conducted comprehensive experiments based on real data captured by AFC systems in Singapore and Taipei to compare the performance of \solution\ and its competitors, including both commercial services and academic contributions. Consistent as our expectation, \solution\ has demonstrated a much better performance in terms of both accuracy and efficiency. In the near future, we plan to extend \solution\ to predict the commuting flows of each individual stations inside a metro network.

\section*{Acknowledgments}
This research is supported by the National Research Foundation, Prime Minister's Office, Singapore under its International Research Centres in Singapore Funding Initiative. The authors would like to thank Land Transport Authority (LTA) for providing the data. However, we declare that all the findings shared in this paper represent the opinions of the authors but not LTA. 

\bibliographystyle{ACM-Reference-Format}
\bibliography{bib}

\end{document}